\documentclass[a4paper, 11pt]{article}

\usepackage[latin1]{inputenc}
\usepackage{lmodern}
\usepackage[T1]{fontenc}
\usepackage{graphicx}
\usepackage[english]{babel}

\usepackage{upgreek}
\newcommand{\smb}{{Sam$\upbeta$ada}}
\usepackage{subcaption}
\usepackage{csquotes}
\usepackage{array}
\usepackage{tabularx}
\usepackage{amsmath}
\usepackage{changepage}
\usepackage{multirow}
\usepackage[a4paper]{geometry}
\usepackage{natbib}

\usepackage{authblk}

\usepackage[setpagesize=false]{hyperref}

\usepackage[left,displaymath, mathlines]{lineno}
\author[1]{Sylvie Stucki}
\author[2]{Pablo Orozco-terWengel}
\author[2]{Michael W. Bruford}
\author[3]{Licia Colli}
\author[4]{Charles Masembe}
\author[3,5]{Riccardo Negrini}
\author[5,6]{Pierre Taberlet}
\author[1,\footnote{to whom correspondence should be addressed}]{St\'{e}phane Joost}
\author[8]{the NEXTGEN Consortium}

\newcommand{\labostyle}[1]{ {\footnotesize #1} }

\affil[1]{\labostyle{Laboratory of Geographic Information Systems (LASIG), School of Architecture, Civil and Environmental Engineering (ENAC), Ecole Polytechnique F\'{e}d\'{e}rale de Lausanne (EPFL), 1015 Lausanne, Switzerland}}
\affil[2]{\labostyle{School of Biosciences, Sir Martin Evans Building, Cardiff University, Cardiff, UK}}
\affil[3]{\labostyle{Istituto di Zootecnica and BioDNA - Centro di Ricerca sulla Biodiversit\`{a} e sul DNA Antico, Universit\`{a} Cattolica del S. Cuore, via E. Parmense 84, 29100 Piacenza, Italy}}
\affil[4]{\labostyle{Makerere University, Institute of Environment and Natural Resources, Molecular Biology Laboratory, Kampala, Uganda}}
\affil[5]{\labostyle{Associazione Italiana Allevatori, Roma, Italy}}
\affil[6]{\labostyle{CNRS, Laboratoire d'Ecologie Alpine, Grenoble, France}}
\affil[7]{\labostyle{Univ. Grenoble Alpes, Laboratoire d'Ecologie Alpine, Grenoble, France}}
\affil[8]{\labostyle{\href{http://nextgen.epfl.ch}{http://nextgen.epfl.ch}}}

\title{High performance computation of landscape genomic models integrating local indices of spatial association}

\date{}

\begin{document}

\maketitle

\begin{abstract}
Since its introduction, landscape genomics has developed quickly with the increasing availability of both molecular and topo-climatic data. 
The current challenges of the field mainly involve processing large numbers of models and disentangling selection from demography. 
Several methods address the latter, either by estimating a neutral model from population structure or by inferring simultaneously environmental and demographic effects. 
Here we present \smb, an integrated approach to study signatures of local adaptation, providing rapid processing of whole genome data and enabling assessment of spatial association using molecular markers.
Specifically, candidate loci to adaptation are identified by automatically assessing genome-environment associations.
In complement, measuring the Local Indicators of Spatial Association (LISA) for these candidate loci allows to detect whether similar genotypes tend to gather in space, which constitutes a useful indication of the possible kinship relationship between individuals.
In this paper, we also analyze SNP data from Ugandan cattle to detect signatures of local adaptation with \smb, BayEnv, LFMM and an outlier method (FDIST approach in Arlequin) and compare their results. 
\smb\ is an open source software for Windows, Linux and MacOS X available at \url{http://lasig.epfl.ch/sambada}.
\end{abstract}


\section{Introduction}

In the late 1970s, \citet{mitton:1977} first had the idea to correlate the frequency of alleles with an environmental variable to look for signatures of selection in ponderosa pine.
Thirty years later, \citeauthor{joost:2007}'s \citeyearpar{joost:2007, joost:2008a} developed the same concept in order to allow parallel processing of large numbers of logistic regressions.
Since then, no noticeable progress was observed in the development of the correlative approaches used in the emerging field of landscape genomics until recently.
During this period correlative approaches were used in parallel with population genetics outlier-detection methods \citep[e.g.][]{beaumont:1996, vitalis:2003, foll:2008} as cross-validation \citep[e.g.][]{jones:2013, henry:2013} to detect signatures of selection \citep[see a review in][]{vitti:2013}. 
However, while such methods are still in vogue \citep[e.g.][]{colli:2014, lv:2014}, there has been a revival in the interest of developing new statistical approaches for the emerging field of landscape genomics \citep[e.g.][]{coop:2010, gunther:2013, frichot:2013, guillot:2014}. 
For example, BayEnv \citep{gunther:2013} implements a Bayesian method to compute correlations between allele frequencies and ecological variables taking into account differences in sample sizes and shared demographic history. 
LFMM \citep{frichot:2013} estimates the influence of population structure on allele frequencies by introducing unobserved variables as latent factors. 
Finally, SGLMM \citep{guillot:2014} uses a spatially-explicit computational framework including a random effect to quantify the correlation between genotypes and environmental variables. 
Yet, important functions are still lacking such as high performance computing capacity to process whole genome data, and the integration of spatial statistics to support a distinction between selection and demographic signals. 
Here we present the software \smb, which aims at filling these gaps by offering an open source multivariate analysis framework to detect signatures of selection. 
\smb's use is illustrated with a case study dedicated to the detection of potentially adaptive loci in 813 \textit{Bos taurus} and \textit{Bos indicus} individuals in Uganda genotyped for $\sim$~40,000 SNP. Lastly, \smb's performance is described with respect to other state of the art software to detect signatures of selection.

\section{New Approaches}

\smb\ uses logistic regressions to model the probability of presence of an allelic variant in a polymorphic marker given the environmental conditions of the sampling locations \citep{joost:2007}.
Since each of the states of a given character is considered independently (i.e.\ as binary presence/absence in each sample), \smb\ can handle many types of molecular data (e.g.\ SNPs, indels, copy number variants and haplotypes), provided the user formats the input. 
Specifically, biallelic SNPs are recoded as three distinct genotypes. 
A maximum likelihood approach is used to fit the models \citep{dobson:2008}.

In the univariate case, each model for a given genotype is compared to a constant model, where the probability of presence of the genotype is the same at each location and is equal to the frequency of the genotype. 
Significance is assessed with both log-likelihood ratio ($G$) and Wald tests \citep{dobson:2008}. 
Bonferroni correction is applied for multiple comparisons. 
In order to avoid numerous computations of $p$-values, the significance threshold $\alpha$ is converted to a minimum score threshold. 
$G$ and Wald scores are used to compare models rather than Akaike or Bayesian information criterion in order to automate model selection.

In comparison to MatSAM \citep{joost:2008a}, \smb\ proposes several improvements: faster processing (see \nameref{implementation} in \nameref{material-methods}), multivariate analysis and measures of spatial autocorrelation.

\subsection{Multivariate analysis} 

Contrary to the univariate approach, the multivariate approach uses several environmental variables to model the presence of each genotype.
The model selection procedure is adapted to assess the significance of multivariate models. 
Both $G$ and Wald tests refer to a null model to build the null hypothesis. 
The current model can be compared to the constant model (the same as in the univariate analysis) using multivariate $\chi^2$ statistics. 
While rejecting the null hypothesis in this configuration indicates that at least one parameter in the model is statistically significant, it does not provide information about which parameter(s) is relevant to the model. 
Therefore \smb\ assesses parameter significance in multivariate models with either a Wald test applied to each parameter separately (except the constant parameter) or with $G$ tests excluding a parameter at the time, that is to say model selection is based on simpler models nested in the current one. 
For the $G$ test, if a model $A$ has $q$ parameters, we define the \emph{parents} of $A$ as the $q$ models with $q-1$ parameters obtained by dropping one parameter from $A$. 
For instance, if $A$ models the occurrence of genotype $X_i$ with 3~environmental variables $E_2$, $E_3$ and $E_5$,
\begin{linenomath}
\begin{equation*}
A = (X_i | E_2, E_3, E_5),
\end{equation*}
\end{linenomath}
then the parents of $A$ are the three models
\begin{linenomath}
\begin{equation*}
(X_i | E_2, E_3), (X_i | E_2, E_5) \text{ and } (X_i | E_3, E_5).
\end{equation*}
\end{linenomath}
The name reflect the fact that $A$ can be obtained by adding a parameter to any of its parents.
The parent of $A$ with the highest log-likelihood is used as the reference model for the significance test. 
This way, the $G$ score is the smallest possible among all parents, thus if the null hypothesis is rejected, it will also be rejected by comparing $A$ with each of its parents.
This method ensures that adding a new parameter leads to a better modelling of the presence of the genotype. 
The overall procedure for fitting and selecting models for each genotype begins with the computation of the constant model. 
Univariate models are built and tested against the constant one, followed by testing bivariate models against their parents, and so forth until the user-defined maximum number of parameters is reached.
Let us notice that this procedure can be applied to any correlative approach relying on significance tests.

Multivariate models allow to take into account pre-existant knowledge provided the data is a continuous variable. 
Specifically, if the population structure was analyzed beforehand and can be represented as a coefficient of membership for each individual, this information can be included in bivariate models.
For models involving both an environmental variable and this coefficient, the selection procedure will assess whether the environmental variable is associated with the genotype while taking into account the possible effect of admixture.
In case there are many ancestral populations, several coefficients may be included in the analysis, leading to a highly conservative detection of selection signatures.

\subsection{Spatial autocorrelation}
Beyond detection of selection signatures, \smb\ quantifies the level of spatial dependence in the distribution of each genotype. 
This measure of spatial autocorrelation refers to similarities or differences between neighbouring individuals that cannot be explained by chance. 
Assessing whether the geographic location has an effect on allele frequency is especially important in landscape genomics since statistical models assume independence between samples. 
Thus if individuals with similar genotypes tend to concentrate in space, spurious correlations may co-occur with specific values of environmental variables. 
On the other hand, spatial independence of data strengthens the confidence in the detections.
It has to be noted that this measure of spatial autocorrelation can be used in complement to any method providing a list of loci possibly under selection.

\smb\ measures the global spatial autocorrelation in the whole dataset with Moran's $I$, as well as the spatial dependency of each point with Local Indicators of Spatial Association (LISA) \citep{moran:1950, anselin:1995}. 
In practice, LISAs are computed by comparing the value of each point with the mean value of its neighbours as defined by a specific weighting scheme based on a kernel function (see supplementary material). 
Both a spatially fixed kernel type relying on distance only, and a varying kernel type considering point density can be used. 
\smb\ includes three fixed kernels (moving window, Gaussian and bisquare) and a varying one (nearest neighbours). 
The sum of LISAs on the whole dataset is proportional to Moran's $I$ \citep{anselin:1995}. 
Significance assessment relies on an empirical distribution of the indices. 
For Moran's $I$, values (genotype occurrences) are permutated among the locations of individuals in the whole dataset and a pseudo $p$-value is computed as the proportion of permutations for which $I$ is equal to - or more extreme (higher for a positive Moran's $I$ or lower for a negative Moran's $I$) - than the observed $I$. 
For LISA, the pseudo $p$-value is separately computed for each point (individual), by keeping the individual of interest fixed and permuting its neighbouring points with the rest of the dataset.

\section{Results}

This study addressed local adaptation of Ugandan cattle, which is composed of two main populations.
Ankole (or Ankole-Watusi) cattle are a Sanga breed (taurine-zebu cross) appeared in the Nile Basin around 2.000 years BC.
They migrated southward and are now found in South-West Uganda, Rwanda and Burundi \citep{ajmone-marsan:2010, ndumu:2008}.
Shorthorn zebus were introduced in East Africa around the VIII\textsuperscript{th} century AD, they later spread as they were less affected than taurine and Sanga cattle to rinderpest, but their susceptibility to trypanosomiasis is presumed to have restrained their progression \citep{ajmone-marsan:2010}.
Shorthorn zebus are now common in North-East Uganda and are interbreeding with Ankole cattle in the center of the country.
These elements match the results of the population structure analysis (see \nameref{pop-structure} in \nameref{material-methods} and figure S1.)

\subsection{Detection of selection signatures}

Four approaches were applied to detect selection signatures among 40,019 SNPs from 804 samples (see \nameref{mol-data} and \nameref{alt-methods-detect-sel} in \nameref{material-methods}). 
The statistical significance threshold for \smb, LFMM and Arlequin was set to 1\% before applying Bonferroni correction. 
For BayEnv, model selection was based on the distribution of Bayes Factors (BF) for neutral loci \citep{coop:2010}. 
Results were analysed separately for each environmental variable and models showing a BF higher than the 1\textsuperscript{st} percentile of the neutral distribution were candidate loci. 
For BayEnv and Arlequin, samples were classified into populations using a threshold of $0.85$ for the higher admixture coefficient, leading to three clusters of 162 Ankole cattle, 8 zebus and 10 cattle from the third population; samples from the fourth population were highly admixed and none satisfied the condition.
This method was preferred to a classification based on sampling locations or phenotypic traits because Ugandan cattle are generally admixed (see figure S1) and these observations could not support a reliable classification.
Since Ugandan cattle is globally constituted of two admixing populations (figure S1), a single coefficient is sufficient to provide an overall view of an individual's ancestry and thus the number of latent factors was set to $1$ in LFMM. 

Using univariate models, \smb\ identified 2,499 SNPs (6.2 \%) potentially subject to selection, BayEnv 1977 (4.9 \%), LFMM 280 (0.7 \%) and Arlequin did not identify any loci as significant. 
The loci detected by \smb\ with the highest $G$ scores were compared among methods in table S3. 
Thirty-six loci were identified by the three correlative methods and three of them were among the most significant models in \smb\ (Table S3). 
These three SNPs occur close to each other in chromosome five. 

\smb's multivariate analysis, which included the population structure as an environmental variable (see \nameref{var-sel-multi} in \nameref{material-methods}), identified 84 significant bivariate models, corresponding to 29 loci. 
In \smb's framework, this means these models provided a significantly more accurate estimation of the genotype's frequency than their univariate parents, while at least one of their parents was also significant. 
Among those, three models that included the \enquote{population structure} variable also had a parent model showing a significant association with this variable. 
Therefore, although the population structure partly explains the distribution of these genotypes, adding an environmental variable provided a significantly more accurate estimation of their distribution  ($p \leq 7.9\cdot 10^{-10}$). 
These models correspond to three loci that were detected by all correlative approaches.

\subsection{Spatial autocorrelation}

\begin{figure*}
\centering
\begin{subfigure}{.46\textwidth}
	\includegraphics[width=\textwidth]{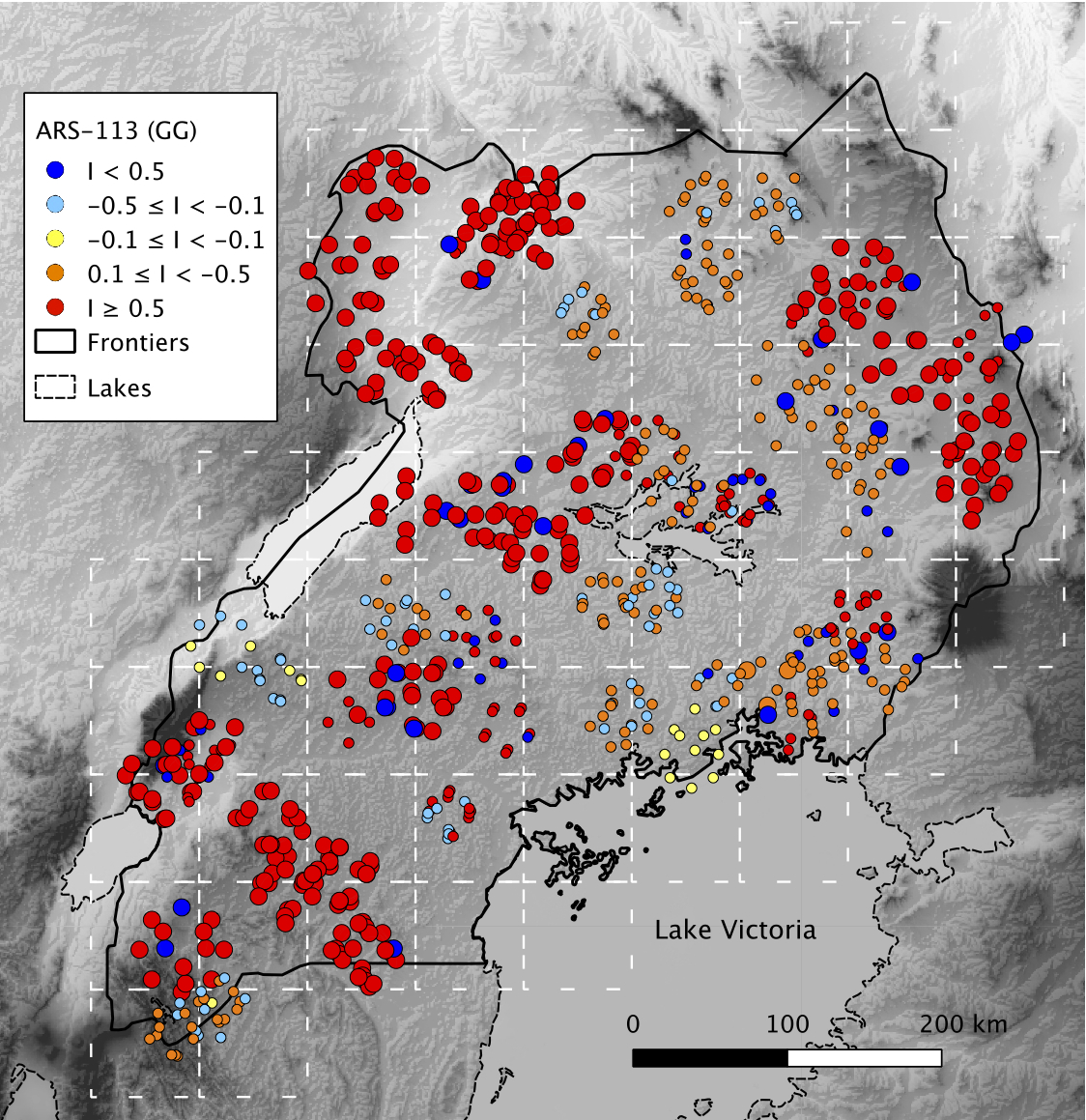}
	\caption{ARS-113 (GG)}
\end{subfigure}%
\hspace{1cm}%
\begin{subfigure}{.46\textwidth}
	\includegraphics[width=\textwidth]{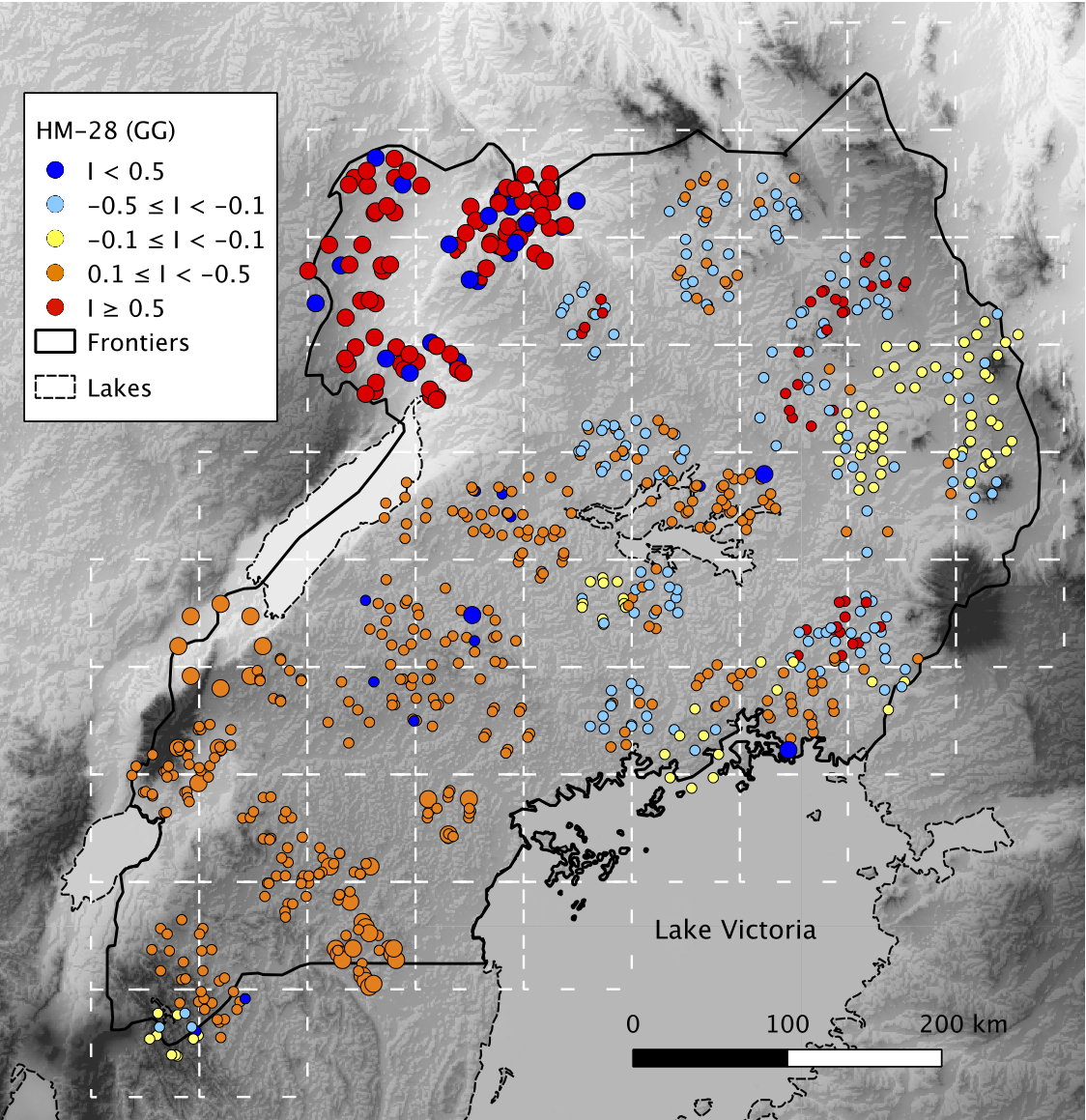}
	\caption{HM-28 (GG)}
\end{subfigure}%
\hfill

\caption{Local indicators of spatial association of markers ARS-113 (allele GG) and HM-28 (allele GG). 
The weighting scheme is based on the 20 nearest neighbours. 
Red points tend to be similar to their neighbours while blue points differ from them. 
Yellow points are independent from their neighbourhood. 
Small points indicate non-significant values ($p>0.001$). 
The map in the background represents the relief, the darker the shade, the higher the altitude.
Samples coming from the same farm have been spread on a circle around their actual location.
}
\label{fig:lisa}
\end{figure*}

Global and local indicators of spatial autocorrelation were computed for two genotypes with a weighting scheme based on the 20 nearest neighbours: ARS-BFGL-NGS-113888 (hereon ARS-113) (allele GG), which was detected by  \smb\ with the highest $G$ score and was also detected by BayEnv, was compared with Hapmap28985-BTA-73836 (here on HM-28) (allele GG), which was detected by  \smb, BayEnv and LFMM. 
Logistic models significantly associated both genotypes with isothermality, which measures the stability of temperature during the year. 
Figure \ref{fig:lisa} shows local indices of spatial autocorrelation for these 2 genotypes. 
On the one hand, ARS-113 (GG) was positively autocorrelated for the majority of points and the indicator was significant for half of them. 
Thus the distribution of this marker shows spatial dependence, non-significant associations were found at the edge of Lake Victoria and in a corridor in the North of the Lake with some occurrences in the West of Uganda. 
This widespread pattern of spatial autocorrelation could originate from the underlying population structure, since Ankole cattle are more common in the South-West while zebus are more common in the North-East of the country. 
Thus the correlation detected by logistic regressions between ARS-113 (GG) and environmental variables could be spurious and due to demographic factors. 
On the other hand, the local indicators of spatial association of HM-28 (GG) showed lower values in general and were only significant in the North-West of Uganda. 
This particular region also showed the lowest values of isothermality in Uganda, i.e.\ a high variability of temperatures. 
The low value of spatial autocorrelation for HM-28 (GG) implies that the distribution of this genotype was mostly independent from the location and this supports a possible adaptive origin of the observed correlation between HM-28 (GG) and isothermality with logistic models. 
This correlation between HM-28 (GG) and isothermality also appeared with bivariate LISAs, where the presence of the genotype was compared to the mean value of isothermality among neighbouring points (not shown).

\section{Discussion\label{discussion}}

The key features of  \smb\ are the multivariate modelling and the measure of spatial autocorrelation. 
Both can support the interpretation of results in case the dataset features population structure. 
Bivariate models may indeed include the global ancestry coefficients provided by a preliminary analysis. 
This setup can detect which loci are correlated with the environment while taking demography into account. 
Additionally, the introduction of measurements of spatial autocorrelation into these analyses integrates spatial statistics with landscape genomics. 
Contrary to most current and non-spatial approaches \citep[e.g.][]{frichot:2013, coop:2010}, \smb\ allows the determination of whether the observed data reflects independent samples, a requirement of the underlying statistical model. 
Measuring spatial autocorrelation assesses whether the occurrence of a genotype is related to its frequency in the surrounding locations. 
More specifically, local indices of spatial autocorrelation allow the mapping of areas prone to spatial dependency. 
On the basis of the present analysis, using spatial statistics in conjunction with correlative models may lower the risk of false positives due to population structure in landscape genomics.

In the present study, \smb\ detected the highest number of SNPs as potentially subject to selection among the four approaches. 
However when comparing the positions of these SNPs, 1,029 of them were less than 100,000 base pairs apart from another detected locus, thus some of these detections might refer to the same signature of selection. 
\smb's results partially match with those of BayEnv with 435 common SNPs (i.e.\ 22\% of BayEnv's detections). 
Concerning the third correlative approach, LFMM is more conservative than \smb\ but the correspondence is better since 154 loci (out of 280, i.e.\ 55\% of LFMM's detections) are detected by both methods. 
Moreover, 25 SNPs detected by LFMM only are less than 100,000 base pairs apart from a locus detected by \smb, potentially identifying the same selection signature. 
The order of detections differed between the two methods, as the most significant loci detected by \smb\ are ignored by LFMM. 
Lastly, Arlequin's best results involved 17 SNPs with $p$-values lower than $10^{-4}$ (significance threshold: $\alpha=2.5 \cdot 10^{-7}$), out of which 2 were common with \smb\ and 16 were common with BayEnv. 
This result suggests that population-based methods, whether using outliers or environmental correlations, tend to detect the same selection signatures. 
On the one hand, \smb's detection rate may indicate the occurrence of some false positives due to population structure; on the other hand, the discrepancy between the results may indicate that the more conservative approaches have some false negatives. 
Comparing the results in the light of spatial dependence gives information about the differences between \smb's and LFMM's detections. 
Maps of local spatial autocorrelation for ARS-113 (GG) and HM-28 (GG) illustrated a general trend: 
LFMM discarded SNPs showing significant local spatial autocorrelation for a large proportion of the sampling locations, while \smb\ detected them. 
Thus measuring local autocorrelation of candidate genotypes may help distinguishing between the effects of local adaptation and those of population structure among \smb's detections. 

\defcitealias{wri:2010}{MAAIF \textit{et al.}, 2010}

Regarding common detections, the three SNPs identified by \smb\ when population structure was included as a covariate were among the common detections of correlatives approaches. 
Thus pre-existent knowledge on demography may be built on to refine correlation-based detections of selection signatures.  
One possible approach could consist of computing population structure and then including one variable summarising this structure in the constant model used by \smb. 
In this way, only genotypes showing a significant correlation with the environment while taking the population structure into account would be detected. 
Concerning the biological function of the common detections, these three loci are located on chromosome 5, near the gene POLR3B whose mouse counterpart is involved in limiting infection by intracellular bacteria and DNA viruses (UniProt, \href{www.uniprot.org}{www.uniprot.org}). 
Moreover, genotype HM-28 (GG) shows spatial autocorrelation in the North-Western part of Uganda and this area overlaps with one of those where the higher load of tse-tse fly (\textit{Glossina} spp.) occur in the country (\citet{abila:2008}; \citetalias{wri:2010}). 
Hence the risk of cattle trypanosomiasis is high in this region and the detected mutations may be involved in parasite resistance.

The increasing availability of large molecular datasets raises challenges regarding their analysis. 
Correlative approaches in landscape genomics enable fast detection of candidate loci to local adaptation. 
However these methods must take into account the effect of population structure \citep{frichot:2013, joost:2013, de-mita:2013}. 
Limited dispersal of individuals leads to spatial autocorrelation of marker frequencies, which may cause spurious correlations with the environment. 
\smb\ addresses the first topic by detecting rapidly selection signatures and the second one by measuring the level of spatial autocorrelation for candidate loci. The next methodological step involves developing spatially-explicit models that directly include autocorrelation. 
\citet{guillot:2014} provide such a model, however the current R-based implementation does not enable whole-genome analysis. 
Alternatively Geographically Weighted Regressions (GWR) measure the spatial stationarity of regression coefficients by fitting a distinct model for each sampling location. 
The number of neighbouring points considered for each sampling location is given by the weighting scheme. 
These models allow some \enquote{local} coefficients to differ between sampling points while some \enquote{global} coefficients are common to all points \citep{fotheringham:2002, joost:2013}. 
Thus GWR enables building a null model where the constant term may vary in space and then refining it by adding a global environmental effect for all locations. 
Comparing these two models would enable an assessment of whether the global environmental effect is needed to describe the distribution of the genotype. 
The key advantage of allowing the constant term to vary in space is to take spatial autocorrelation into account in the models. 
This way, GWR allows an investigation of the spatial behaviour of loci showing selection signature with standard logistic regressions and may help to distinguish between local adaptation and population structure in landscape genomics. However GWR models require a fine-tuning of the weighting scheme from the user, which restrains their application to very large datasets.

The recent availability of whole genome sequence (WGS) data also raises issues regarding the statistical assessment of multiple comparisons.
Indeed, while many individuals and few genetic markers were available ten years ago, the current high costs of WGS limit the number of sequenced samples. 
Therefore standard procedures for multiple comparisons, such as the Bonferroni correction, are over-conservative and may lead to discard some adaptive loci.
In this context, alternatives procedures focus on controlling the ratio of false positives in a set of significant results.
Among them, \citeauthor{storey:2003}'s false discovery rate \citeyearpar{storey:2003} was especially designed for large molecular datasets and suits any detection method relying on significance tests. 
Its implementation in \smb\ is ongoing.

Computation time is critical when processing large datasets. 
In this context, \smb\ is able to swiftly analyse high-density SNP-chips and variants from whole-genome sequencing (e.g.\ the case study presented in here is analysed within 69 minutes for univariates models alone and 8.5 hours for both univariate and bivariate models using a single core at 3.1 GHz on a desktop computer with 8 GB of RAM). 
When considering single-process computations, \smb\ is approximately 4.5 times quicker than LFMM and 30 times than BayEnv (see table S2). 
Both \smb\ and LFMM enable parallelised processing. 
\smb's processing speed, combined with its ability to analyse the spatial autocorrelation in molecular data and to incorporate prior knowledge on population structure, suits a wide range of applications, especially those involving whole genome sequence data.

\section{Material and Methods\label{material-methods}}

\subsection{Ugandan cattle}

\subsubsection{Sampling design\label{sampling-design}}

Sampling was designed to cover the whole country, including each eco-geographic region, and to obtain a homogeneous distribution of individuals across the country. 
A regular grid made of 51 cells of 70 x 70 km was produced to this end. 
The field campaign took place between March 2011 and January 2012.
On average, four farms were visited in each cell and four unrelated individuals were selected from each farm, for a total of 917 biological samples retrieved from 202 farms. 
Recorded information also included the location of the farm, the name of the breed, a picture and morphological information on each individual. 
These elements were stored in a database accessible through a Web interface, enabling real-time monitoring of the sampling campaign.

\subsubsection{Molecular data\label{mol-data}}

Out of the 917 individuals, 813 samples were genotyped with a medium-density SNP chip (54,609 SNPs, BovineSNP50 BeadChip, Illumina Inc., San Diego, CA). 
Only markers located on the autosomal chromosomes were considered in the analyses. 
The dataset was filtered with PLINK \citep{purcell:2007} with a call rate set to 95\% for both individuals and SNPs, and a minimum allele frequency (MAF) set to 1\%. 
The resulting dataset contains 804 samples and 40,019 SNPs.

\subsubsection{Population structure\label{pop-structure}}

Population structure was analysed with Admixture \citep{alexander:2009}, which estimates maximum likelihood of individual ancestries from multilocus SNP genotype datasets. 
This approach assumes that samples descend from a predefined number of ancestor populations that are mixing. 
Admixture estimates both the fraction of each sample coming from each population and the marker frequencies in these populations. 
The optimal number of populations is assessed by a $k$-fold cross-validation procedure. 

The best partition of the dataset consisted of four populations, although the vast majority of the samples were allocated to two clusters (almost 96\%) on the basis of the ancestry coefficients (Figure S1). 
Mapping these coefficients revealed the two clusters (340 and 431 individuals out of 804) occurred in the South-West and North-East of Uganda respectively. 
Using pictures of sampled individuals, the first cluster was identified as Ankole cattle and the second one as zebu. 
The remaining two clusters (32 animals) possibly represent introgression from allochthonous gene pools.
These results were used to define the parameters needed by each method to detect selection signatures.

\subsubsection{Environmental data\label{env-data}}

The geographical coordinates of the individuals sampled enabled the characterisation of their habitat with the help of the WorldClim dataset containing monthly values of precipitation, minimum, mean and maximum temperature as well as 19 derived variables, at 1km resolution \citep{hijmans:2005}. 
This dataset provides appropriate data as it consists of representative climate information collected during 30 years \citep[WMO standard climate normal, ][]{arguez:2010} and its high resolution suited the scale of our study.
These environmental variables were originally stored in four tiles (portions of map) which were pasted using the Geospatial Data Abstraction Library \citep{gdal-development-team:2013} and a customized Python script. 
The topography is described by the 90m resolution SRTM3 (Shuttle Radar Topography Mission) digital elevation model \citep{farr:2007}. 
SAGA GIS (\href{www.saga-gis.org}{www.saga-gis.org}) was used to paste the 36 tiles covering the country and to derive slope and orientation from the altitude. 
Longitude and latitude were also included as a proxy of population structure. 
Finally the values of the 72 environmental variables were extracted for each animal using the \enquote{Point Sampling Tool} extension (\href{http://hub.qgis.org/projects/pointsamplingtool}{http://hub.qgis.org/projects/pointsamplingtool}) in QuantumGIS (\href{www.qgis.org}{www.qgis.org}).

\paragraph{Variable selection for univariate models\label{var-sel-uni}} 
Considering all environmental variables in the computation of the multiple logistic regressions would have provided a comprehensive analysis with a low risk of missing detections. 
Nonetheless some variables are highly correlated; this implies dependence between models and increases the variance of parameters in multivariate analyses. 
Thus we used the Variance Inflation Factor (VIF) to control for multicollinearity \citep{dobson:2008}. 
A maximum VIF of 5 corresponds to a coefficient of correlation of 0.9 between pairs of variables. 
The number of variables was reduced iteratively by randomly removing one of the two most highly correlated variables until the maximum correlation was lower than the threshold (0.9). 
This procedure led to a set of 23 environmental variables that were used for univariate landscape genomic analysis (table S1). 

\paragraph{Variable selection for multivariate models\label{var-sel-multi}} 
Bivariate models were also processed with \smb\ to assess the effect of a combination of predictors, and to take the population structure into account. 
This information was derived from the analysis of individual ancestries (see \nameref{pop-structure}). 
Since there were two main populations of Ugandan cattle, the associated coefficients of membership were generally complementary and a single coefficient was sufficient to represent the origin of each individual.
Thus a new variable \enquote{population structure} was defined as the coefficient of membership of each individual to the population Ankole.
This variable was used to represent the ancestry of each sample in \smb's analysis.
It was added to the set of 23 variables and the correlation-based variable selection method was reapplied to limit the VIF to 5. 
On this basis, fifteen environmental variables were considered for \smb\ bivariate analysis (table S1).

\subsection{\smb\ implementation\label{implementation}}

\smb\ is a standalone application written in C++. 
The application was developed using the Scythe Statistical Library \citep{pemstein:2011} for matrix computation and probability distributions, which was also chosen for its straightforward application programming interface (API). 
\smb\ is distributed under an open source GNU General Public License license in order to ease its use for research and teaching. 

\subsubsection{Desktop and High Performance Computing}

Two complementary versions of the software were developed: a desktop option-rich program well suited to small to medium-sized datasets, and a High Performance Computing version dedicated to large datasets.

\paragraph{Desktop version (\smb)}

\smb\ includes multivariate analysis and spatial autocorrelation. 
Many options are provided to facilitate the formatting of the data and to customise the analysis. 
For instance, the significance of models is assessed during the analysis and non-significant associations can be discarded. 
Moreover models can be sorted according to their scores before writing the results in order to make it possible to directly be in a position to interpret them. 
All results presented in this paper were processed with \smb\ Desktop.

\paragraph{Parallel computing version (CoreSAM)}

When processing large datasets, primary analysis usually focuses on univariate models. 
Multivariate models and spatial autocorrelation may be considered as a second step, but are too computationally intensive to be applied to the whole dataset. 
In order to speed-up the process, CoreSAM is a light version of \smb, written in C, which focuses on univariate analysis. 
Compared with \smb, fewer options are available, but the computation is up to seven times faster.

Combining \smb\ and CoreSAM, large datasets may be analysed by steps: 
Univariate logistic models identify candidate loci exhibiting selection signatures. 
These loci may be then investigated in the light of spatial autocorrelation measures and multivariate models. 
The former may point out whether the observed correlation is due to similarities between neighbours, while the latter allows including the population structure, if any, in the model in order to assess whether the environmental variable provides supplementary information on the marker frequency when taking the demography into account.

\subsubsection{Modules}

\smb\ includes several modules that enhance interfacing with other programs.

\paragraph{Geovisualization of spatial statistics}

\smb\ provides an option to save the spatial autocorrelation results as a shapefile (.shp), a common format for storing vector information in Geographic Information Systems (GIS). 
This feature relies on the shplib open source library (\url{http://shape-lib.maptools.org}).

\paragraph{Recoding molecular data}
 
\smb\ is distributed with a utility for recoding molecular data into binary information. Currently RecodePlink handles ped/map files, a format for SNP data used by PLINK \citep{purcell:2007}.  

\paragraph{Supervision}

For very large molecular datasets, \smb\ provides a module to share workload between computers. 
Supervision splits the input data in several files that can be processed separately, even on heterogeneous computers. 
Lastly, Supervision merges the results to provide the same output as if the whole dataset had been processed at once.

\subsection{Alternative methods to detect selection\label{alt-methods-detect-sel}}

For comparison purpose, three alternative approaches to \smb\ were used to detect signatures of selection in Ugandan cattle data. 
Two of these are correlative approaches \citep[BayEnv and Latent Factor Mixed Models,][]{coop:2010, frichot:2013}, while the third is an outlier-detection population genetics approach \citep{beaumont:1996} included in Arlequin 3.5 \citep{excoffier:2010}.
These methods consider allele counts whereas \smb\ recodes them into genotypes.

\subsubsection{BayEnv}
BayEnv uses a Bayesian approach to detect candidate SNPs under selection while accounting for the inherent correlation in allele frequencies across populations due to shared demographic history \citep{coop:2010}. 
BayEnv first uses a set of neutral loci to build a null model robust to demographic history, against which an alternative model including an environmental variable is compared. 
Markers exhibiting the highest Bayes factors are potentially subject to selection. 
In this study the set of neutral loci was chosen at random among loci identified as neutral by \smb.

\subsubsection{Latent Factor Mixed Models}
LFMM is a Bayesian individual-based approach to detect selection in landscape genomics \citep{frichot:2013}. 
Population structure is added into the model via unobserved variables. 
Thus the significance of the association between genome and environment can be assessed while taking into account the effect of the population structure. 
The number of latent factors (unobserved variables) must be specified for the analysis. 
Although this number is related to the population structure, it is usually lower than the number of populations (Frichot personal communication).

\subsubsection{Outlier approach}
Arlequin is a comprehensive software for population genetics analyses \citep{excoffier:2010} that includes an outlier-based method to detect signatures of selection \citep{beaumont:1996}. 
This approach assumes an island model, where individuals are sampled in distinct populations that exchange migrants. 
Each locus is characterised by the fixation index $F_{ST}$ \citep{wright:1949} and its heterozygosity. 
Neutral coalescent simulations are used to estimate the distribution of $F_{ST}$ conditional on heterozygosity. 
Loci exhibiting extreme values of $F_{ST}$ are candidate targets of selection.

\subsection{Resources\label{resources}}

\subsubsection{Software availability}
\smb\ is an open source software written in C++ available at \url{http://lasig.epfl.ch/sambada} (under the license GNU GPL~3). 
Compiled versions are provided for Windows, Linux and MacOS X.

\subsubsection{Data availability}
NextGen data are described at \url{http://projects.ensembl.org/nextgen/}.
Ugandan cattle SNP data are available at 
\url{ftp://ftp.ebi.ac.uk/pub/databases/nextgen/bos/variants/chip_array/}
in PLINK format (files UGBT.bovineSNP50.UMD3\_1.20140307.[ped/map].gz)
with the following data policy
\url{ftp://ftp.ebi.ac.uk/pub/databases/nextgen/documentation/README_data_use_policy}.

\section{Supplementary material}

Supplementary material, figure S1, and tables S1--S3 are available at Molecular Biology and Evolution online (\url{http:// www.mbe.oxfordjournals.org/}).

\section{Acknowledgement}
\paragraph{Funding:} This research was funded by EU FP7 project NextGen (Grant KBBE-2009-1-1-03).
 
The authors would like to thank Sergio Rey for his advice on assessing the significance of LISA, Stephan Morgenthaler for fruitful discussions on assessing the significance of multivariate logistic models and Gilles Guillot for providing them with SGLMM for testing purposes.

\bibliographystyle{natbib}
\bibliography{sambada}

\end{document}


\title{High performance computation of landscape genomic models integrating local indices of spatial association\\ \emph{Supplementary material}}
\date{}
\maketitle

\renewcommand\thefigure{S\arabic{figure}}    
\renewcommand\thetable{S\arabic{table}}


\section{Spatial autocorrelation}

Several indices are available for measuring the global spatial autocorrelation in a dataset. 
\smb\ uses the Moran's $I$ \citep{moran:1950}, defined as follows:
	\begin{equation}\label{eq:i-moran-global}
			I = \frac{n}{S_0} \frac{\sum_{i=1}^n  \sum_{j=1}^n w_{ij} (y_i - \bar{y})  (y_j - \bar{y}) }{\sum_{k=1}^n (y_k - \bar{y})^2} 
			   = \frac{n}{S_0} \frac{\sum_{i=1}^n  \sum_{j=1}^n w_{ij} z_i z_j }{\sum_{k=1}^n z_k^2}
	\end{equation}
with

	\begin{tabular}{ll}
	$n$ & number of sampling points;\\
	$w_{ij}$ & weight of point j in the neighbourhood of i, defined by the spatial kernel;\\
	$S_0$ & sum of all weights  $\left(S_0 = \sum_{i=1}^n  \sum_{j=1}^n w_{ij} \right)$; \\
	$y_i$, $y_j$ & 	values for points i and j;\\
	$\bar{y}$ & mean value; \\
	$z_i$, $z_j$ & deviations from the mean.\
	\end{tabular}

Please note that the weight matrix is usually normalized per line $\left(\sum_{j=1}^n w_{ij} = 1 \right)$, so that $S_0$ is equal to $n$. Thus equation \ref{eq:i-moran-global} can be rewritten as:
	\begin{equation}
			I   =   \frac{\sum_{i=1}^n  \sum_{j=1}^n w'_{ij} z_i z_j }{\sum_{k=1}^n z_k^2}
	\end{equation}
where $w'_{ij}$ is the normalized weight of point j in the neighbourhood of i.

Local indices of spatial association \citep[LISA,][]{anselin:1995} measure the local association between the value of a point and the neighbouring points. 
\smb\ computes a local variant of the Moran's $I$ for each point $i$:

\begin{align}
	I_i 	&= \frac{n-1}{\sum_{k=1}^n z_k^2}  \quad \left( z_i \sum_{j=1}^n w'_{ij} z_j\right)
\end{align}

LISAs are defined in such a way that their sum over all points is proportional to a global measure of spatial autocorrelation \citep[here the Moran's $I$,][]{anselin:1995}:

\begin{align}
	I 	&= \frac{1}{n-1} \sum_{i=1}^n I_i
\end{align}

\clearpage

\section{Environmental variables}

\begin{table}[h!]


\begin{tabular}{| L{2cm} | L{5.1cm}  | L{3cm} | C{2cm} | C{2cm} | }
\hline
Variable  	&	Description			&	Data source	&	Used for univariate analysis 	&	Used for bivariate analysis	\\
\hline
alt\_SRTM&	Altitude [m] 			&	SRTM3	&	X	&		\\
\hline
aspect	&	Orientation of the relief [°]	&	Derived from SRTM3	&	X	&	X	\\
\hline
BIO2 	&	Mean Diurnal Range	&	\multirow{14}{3cm}{WorldClim}	&	\multirow{3}{2cm}{\centering X}	&	\multirow{3}{2cm}{\centering X}	\\
		&	(Mean of monthly (max temp - min temp))		&		&		&		\\
\cline{1-2}\cline{4-5}
BIO3  	&	Isothermality (BIO2/BIO7) (* 100)	&		&	X	&	X	\\
\cline{1-2}\cline{4-5}
BIO7  	&	Temperature Annual Range (max temp - min temp)	&		&	X	&		\\
\cline{1-2}\cline{4-5}
BIO9  	&	Mean Temperature of Driest Quarter	&		&	X	&		\\
\cline{1-2}\cline{4-5}
BIO12  	&	Annual Precipitation	&		&	X	&	X	\\
\cline{1-2}\cline{4-5}
BIO15  	&	Precipitation Seasonality (Coefficient of Variation)	&		&	X	&	X	\\
\cline{1-2}\cline{4-5}
BIO18  	&	Precipitation of Warmest Quarter	&		&	X	&	X	\\
\hline
latitude	&	Latitude	&	\multirow{2}{3cm}{Sampling measurements}	&	X	&	X	\\
\cline{1-2}\cline{4-5}
longitude 	&	Longitude 	&		&	X	&	X	\\
\hline
prec2	&	Precipitation in February	&	\multirow{9}{3cm}{WorldClim}	&	X	&		\\
\cline{1-2}\cline{4-5}
prec3	&	Precipitation in March	&		&	X	&		\\
\cline{1-2}\cline{4-5}
prec4	&	Precipitation in April	&		&	X	&	X	\\
\cline{1-2}\cline{4-5}
prec5	&	Precipitation in May	&		&	X	&	X	\\
\cline{1-2}\cline{4-5}
prec6	&	Precipitation in June	&		&	X	&		\\
\cline{1-2}\cline{4-5}
prec7	&	Precipitation in July	&		&	X	&		\\
\cline{1-2}\cline{4-5}
prec9	&	Precipitation in September	&		&	X	&		\\
\cline{1-2}\cline{4-5}
prec10	&	Precipitation in October 	&		&	X	&	X	\\
\cline{1-2}\cline{4-5}
prec11	&	Precipitation in November	&		&	X	&	X	\\
\hline
slope	&	Slope of the relief [\%] 	&	Derived from SRTM3	&	X	&	X	\\
\hline
tmin10	&	Minimal temperature in October	&	\multirow{2}{3cm}{WorldClim}	&	X	&	X	\\
\cline{1-2}\cline{4-5}
tmax10	&	Maximal temperature in October	&		&	X	&		\\
\hline
Ankole	&	Coefficient of ancestry to the population Ankole	&	Analysis with Admixture	&		&	X	\\
\hline
\multicolumn{3}{|l|}{\emph{Number of variables}}				&	23	&	15	\\
\hline
\end{tabular}
\caption{Environmental variables used to detect selection signatures with correlative approaches.  Univariate analyses were performed with Sambada, BayEnv and LFMM and bivariate analyses with Sambada}
\label{tab:variables}
\end{table}
\restoregeometry

\section{Population structure}

\begin{figure}[h]
\centering

	\includegraphics[width=42pc]{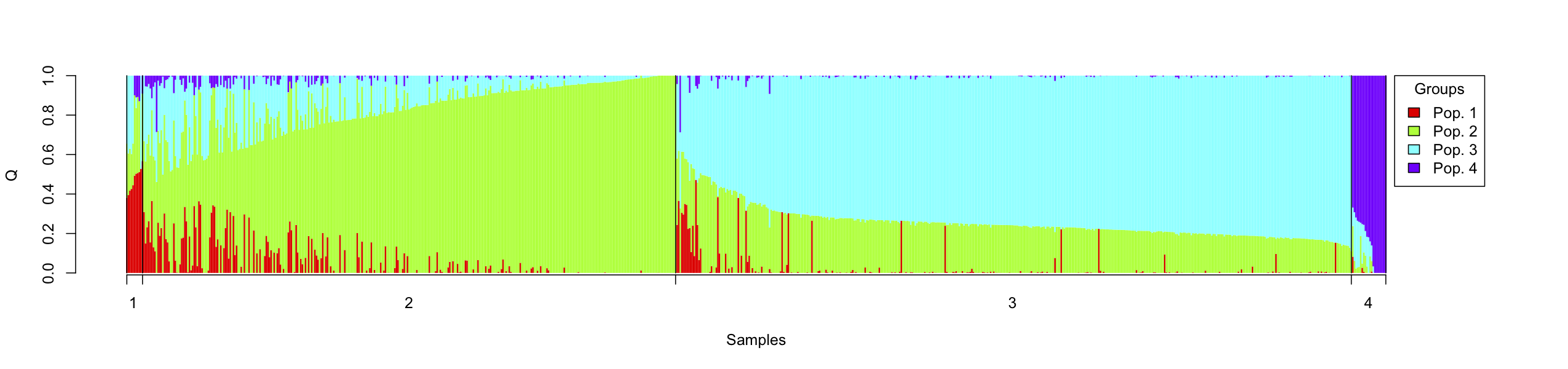}
\hfill

\caption{
Population structure computed with Admixture \citep{alexander:2009}.
Individuals are gathered together by populations, labeled horizontally.
The assignation is based on the highest membership coefficient  $Q_{\text{max}}$ of the sample.
Inside each population, individuals are ranked by increasing (or decreasing) value of  $Q_{\text{max}}$.
}

\end{figure}

\section{Computation time}

\begin{table}[h]
\centering
\begin{tabular}{|l|rr|}
\cline{2-3}
\multicolumn{1}{l|}{}		&	41,215 SNPs & 634,849 SNPs \\					
\multicolumn{1}{l|}{}		&	804 samples	& 102 samples\\
\hline
\smb		&	1.2			& 2.9	\\
\smb\ biv. 	&	8.7			& 18.4	\\
{BayEnv}	& 41.3		& 62.,2	\\
{LFMM}	& 3.2 		& 16.0 \\
{LFMM} (mono)	& 6.1		& 58.1\\
\hline
\end{tabular}	
\caption{Comparison of computation times among methods.
The datasets used in this case include chromosome X. 
The first column refers to the data used in the article (SNPs and individual call rates=5\%, MAF=1\%, including chr. X). 
The second column refers to an unpublished dataset of 102 samples of Ugandan cattle that were chosen among the 917 samples to be genotyped with a high-density SNP chip (BovineHD, Illumina Inc; SNPs and ind. call rates=5\%, MAF=5\%). 
Durations are expressed in hours.  
\enquote{LFMM (mono)} shows the duration if using a single thread. 
Analyses were run on a desktop computer with an 8-core CPU at 3.1 GHz and 8 GB of RAM.
No computation time is available for Arlequin.
}	
\end{table}

\clearpage

\section{Comparison of detections}
\begin{table}[ht]
\centering
 \begin{tabular}{|+r^l*7{^r}|}
  \hline
  & Loci & Chr.  & Pos. [Mbp] & \rotatebox{90}{\smb} & \rotatebox{90}{{BayEnv}} &  \rotatebox{90}{{LFMM}} &  \rotatebox{90}{{Arlequin} } & Detections \\ 
  \hline
  1 & ARS-BFGL-NGS-113888 & 5 & 48.32 & 1 & 1 & 0 & 0 & 2 \\ 
  2 & Hapmap41074-BTA-73520 & 5 & 48.35 & 1 & 1 & 0 & 0 & 2 \\ 
  3 & Hapmap41762-BTA-117570 & 5 & 18.94 & 1 & 1 & 0 & 0 & 2 \\ 
  4 & ARS-BFGL-NGS-46098 & 20 & 2.95 & 1 & 1 & 0 & 0 & 2 \\ 
  5 & Hapmap41813-BTA-27442 & 5 & 49.04 & 1 & 1 & 0 & 0 & 2 \\ 
  6 & BTA-73516-no-rs & 5 & 48.75 & 1 & 1 & 0 & 0 & 2 \\ 
  \rowstyle{\bfseries} 7 & Hapmap28985-BTA-73836 & 5 & 70.34 & 1 & 1 & 1 & 0 & 3 \\ 
  8 & Hapmap31863-BTA-27454 & 5 & 48.99 & 1 & 1 & 0 & 0 & 2 \\ 
   \rowstyle{\bfseries} 9 & ARS-BFGL-NGS-106520 & 5 & 70.20 & 1 & 1 & 1 & 0 & 3 \\ 
  \rowstyle{\bfseries}  10 & BTA-73842-no-rs & 5 & 70.18 & 1 & 1 & 1 & 0 & 3 \\ 
  11 & Hapmap50523-BTA-98407 & 5 & 46.74 & 1 & 1 & 0 & 0 & 2 \\ 
  12 & BTB-01400776 & 20 & 2.70 & 1 & 1 & 0 & 0 & 2 \\ 
  13 & Hapmap23956-BTA-36867 & 15 & 47.20 & 1 & 1 & 0 & 0 & 2 \\ 
  14 & ARS-BFGL-NGS-10586 & 2 & 128.64 & 1 & 1 & 0 & 0 & 2 \\ 
  15 & ARS-BFGL-NGS-43694 & 5 & 49.65 & 1 & 1 & 0 & 0 & 2 \\ 
  16 & BTA-122374-no-rs & 14 & 16.44 & 1 & 1 & 0 & 0 & 2 \\ 
  17 & BTB-01356178 & 20 & 2.49 & 1 & 1 & 0 & 0 & 2 \\ 
   \rowstyle{\bfseries}  18 & ARS-BFGL-NGS-94862 & 11 & 103.53 & 1 & 1 & 1 & 0 & 3 \\ 
  19 & BTA-108359-no-rs & 14 & 16.31 & 1 & 0 & 0 & 0 & 1 \\ 
  20 & ARS-BFGL-NGS-15960 & 5 & 28.02 & 1 & 1 & 0 & 0 & 2 \\ 
  21 & ARS-BFGL-NGS-116294 & 2 & 128.58 & 1 & 1 & 0 & 0 & 2 \\ 
  22 & INRA-566 & 13 & 57.94 & 1 & 0 & 1 & 0 & 2 \\ 
  \rowstyle{\bfseries}  23 & BTA-49720-no-rs & 5 & 69.66 & 1 & 1 & 1 & 0 & 3 \\ 
  24 & ARS-BFGL-NGS-56387 & 13 & 24.36 & 1 & 1 & 0 & 0 & 2 \\ 
  25 & BTA-28185-no-rs & 26 & 22.78 & 1 & 0 & 0 & 0 & 1 \\ 
\hline
\end{tabular}
\caption{List of SNPs detected by \smb\ corresponding to the models with the highest $G$ scores. 
Loci are identified by their name, their chromosome and their position in million base pairs.
The following columns show which method detected them and the last one counts these detections.
Loci in bold type are the commons discoveries of \smb, LFMM and BayEnv.
Local indices of spatial autocorrelation were computed for SNPs on lines 1 and 7.}
\label{tab:comp-detections}
\end{table}

\bibliographystyle{natbib}
\bibliography{sambada}